\documentstyle[epsfig,preprint,aps,eqsecnum,fixes,floats]{revtex}
\tightenlines
\begin {document}

\def\Tc{T_{\rm c}}
\def\kB{k_{\rm B}}

\def\grad{\mbox{\boldmath$\nabla$}}
\def\p{{\bf p}}
\def\q{{\bf q}}
\def\k{{\bf k}}
\def\l{{\bf l}}
\def\P{{\bf P}}
\def\LO{{\rm LO}}
\def\half{{\textstyle{1\over2}}}
\def\f{F}
\def\Sinh{{\rm Sinh}}
\def\Li{{\rm Li}}

\def\Cos{{\rm Cos}}



\title
{$\Tc$ for dilute Bose gases: beyond leading order in $1/N$}

\author {Peter Arnold and Boris Tom\'{a}\v{s}ik}
\address
    {%
    Department of Physics,
    University of Virginia \\
    382 McCormick Road,
    P.O. Box 400714,
    Charlottesville, VA  22904-4714
    }%
\date{May 11, 2000}

\maketitle
\vskip -20pt

\begin {abstract}%
{%
   Baym, Blaizot, and Zinn-Justin have recently used the large $N$
   approximation to calculate the effect of interactions on the
   transition temperature of dilute Bose gases.
   We extend their calculation to next-to-leading-order in $1/N$ and
   find a relatively small correction of -26\% to the leading-order result.
   This suggests that the large $N$ approximation works surprisingly well in
   this application.
}%
\end {abstract}

\thispagestyle{empty}


\section {Introduction}

Second-order phase transitions have universal behavior,
associated with
long wavelength fluctuations, for which critical exponents and other
universal quantities can often be successfully calculated using
renormalization group techniques.  For most such systems, the short
distance physics is hopelessly complicated.
In contrast, the phase transition of a dilute, interacting Bose gas
provides a fascinating example where physics becomes simpler, and
perturbative, at (relatively) small distance scales.
For this system, it should be possible to marry
techniques for treating long-distance critical fluctuations to a
perturbative treatment of short distance physics, and so
compute non-universal characteristics of the phase transition.
A simple example of such a non-universal quantity is the phase
transition temperature $\Tc$, and the effect of interactions on
$\Tc$ has been explored by
several authors \cite{GCL97,HGL99,stoof,schakel,Baym1,holzmann}, with a wide
variety of theoretical results.
In particular, the transition temperature
has recently been calculated by Baym, Blaizot, and Zinn-Justin
\cite{baymN}
in the large $N$ approximation.
For simplicity,
they implicitly focus on the case of Bose gases with a single spin
state, where the low-energy cross-section for atomic collisions can
be parametrized by a single scattering length, $a$.
As will be briefly reviewed below, the problem is first reduced to
a calculation in a three dimensional O(2) scalar field theory at its
critical point.  Replacing that by an O($N$) theory with $N=2$, they
find
\begin {equation}
   \Tc = T_0 \left[1 +
        {8\pi\over 3 \zeta(3/2)^{4/3}} \, a n^{1/3}
        \left[1 + O(N^{-1})\right]
        + O\left((a n^{1/3})^2\right) \right]
\label {eq:TcLO}
\end {equation}
in the dilute limit,
where
$n$ is the number density,%
\footnote{
   For simplicity, we consider a uniform Bose gas, where $n$ is fixed.
   Alternatively, in an arbitrarily  wide harmonic trap,
   $n$ should be interpreted as the actual density
   at the center of the trap at the transition temperature.
}
and $T_0$ is
the transition temperature of a non-interacting Bose gas,
\begin {equation}
   T_0 = {2\pi\hbar^2\over \kB m} \left(n \over \zeta(3/2)\right)^{2/3} .
\label {eq:T0}
\end {equation}
Their result of
$\Delta\Tc/T_0 \equiv (\Tc-T_0)/T_0 \simeq 2.33 \, a n^{1/3}$ is in
good agreement with recent%
\footnote{
   As alluded to earlier, there have 
   been several different theoretical results and simulation results
   obtained by
   various methods
   (e.g. \cite{GCL97,HGL99,stoof,schakel}),
   giving a large range of values for the coefficient of $a n^{1/3}$ in
   $\Delta\Tc/T_0$.
   There has also been some experimental data on the ${}^4$He-Vycor
   system \cite{vycor},
   which superficially seems to fit well an early theoretical
   estimate of Stoof \cite{stoof}, which is
   $\Delta\Tc/T_0 \simeq (16\pi/3)\, \zeta(3/2)^{-4/3} a n^{1/3} \simeq
   4.66 \, a n^{1/3}$.
   However, the detailed
   interpretation of this data
   is unclear.  In that experiment, the Helium atoms are confined to
   an interconnected network of channels in the porous Vycor glass,
   and, for the low-density data that appears to fit Stoof, the interpaticle
   spacing is the same order of magnitude as the widths of the channels.
   Ref.\ \cite{vycor} simply assumes that the system can
   be modeled by a free Bose gas with (i) an effective mass for the atoms
   that is extracted experimentally, but (ii) the same scattering length
   as for bulk Helium, which is moreover taken from theoretical modeling.
   Because of these assumptions, the apparent agreement with
   Ref.\ \cite{stoof} should be treated with caution.
}
numerical simulations \cite{holzmann}
that give $\Delta\Tc/T_0 \approx (2.2\pm0.2) \, a n^{1/3}$.
The result is surprising because it seems to work much better than
the large $N$ expansion of critical exponents.  For example, for
$O(N)$ theory, the susceptibility critical exponent $\gamma$ is
\cite{largeNexp}%
\footnote{
   For Bose gases, a more physical example of a critical exponent is
   $\nu = 1 - 0.540 (2/N) - 0.470 (2/N)^2 + O(N^{-3})$, whose actual
   value is $\nu \simeq 0.67$ for $N=2$.
   The fact that O(2) critical exponents should be identified with
   Bose gas exponents is not completely trivial.  A uniform,
   non-relativistic Bose gas
   is a constrained system: the particle density
   $n$ is fixed.  This constraint causes the critical exponents
   $\tilde x = (\tilde\alpha, \tilde\beta, \tilde\gamma, \tilde\nu)$
   of the actual system to be related \cite{fisher} to the standard
   exponents $x=(\alpha, \beta, \gamma, \nu)$ of the field theory by
   (i) $\tilde \alpha = -\alpha/(1-\alpha)$, and $\tilde x = x/(1-\alpha)$
   for the others, if $\alpha > 0$, or
   (ii) $\tilde x = x$ if $\alpha < 0$.
   The actual value of $\alpha$ for the
   O(2) model is believed to be $-0.007 \pm 0.006$ \cite{alpha}.
   If negative, there is no difference between the exponents;
   if positive, there is in principle a very tiny difference.
   This relation explains, by the way, the difference between mean-field
   theory exponents for the O(2) model (e.g.\ $\alpha = 1/2$) and the
   exponents of a non-interacting Bose gas (e.g.\ $\tilde\alpha = -1$).
}
\begin {eqnarray}
   \gamma &=& 2 - {24\over N\pi^2}
     + {64\over N^2\pi^4}\left({44\over9} - \pi^2\right) + O(N^{-3})
\nonumber\\
   &=& 2 - 1.216 \left(2\over N\right) - 0.818 \left(2\over N\right)^2
      + O(N^{-3}).
\end {eqnarray}
This is not a marvelous expansion for $N=2$, for which the actual
value is $\gamma \simeq 1.32$.

In this paper, we calculate the $O(N^{-1})$ relative correction to
(\ref{eq:TcLO}).  We find
\begin {equation}
   \Tc = T_0 \left[1 +
        {8\pi\over 3 \zeta(3/2)^{4/3}} \, a n^{1/3}
        \left[1 - {0.527198\over N} + O(N^{-2})\right]
        + O\left((a n^{1/3})^2\right) \right] .
\label {eq:TcNLO}
\end {equation}
Setting $N=2$, this is only an 26\% correction to the leading
large $N$ result
for $\Delta\Tc/T_0$.
We now have $\Delta\Tc/T_0 \simeq 1.71\,an^{1/3}$. Though this does
not agree as well with the quoted simulation result, the
moderatley
small size of the correction supports the proposition that the large $N$
expansion works surprisingly well for $\Tc$.

In the remainder of this introduction, we review the
long-distance O(2) effective theory for Bose condensation and then
review the arguments of \cite{Baym1,baymN} about how to calculate
$\Delta\Tc/\Tc$.
In Sec.\ \ref{sec:LO}, we review the
leading-order calculation in large $N$ as done in \cite{baymN}.
In Sec.\ \ref{sec:NLO}, we go on to calculate the next order in $1/N$.
An appendix explains how to calculate some of the basic 3-dimensional
integrals that appear at that order.


\subsection {Review of effective theory}

The basic assumption throughout will be that the average separation
$n^{-1/3}$ of atoms is large compared to the scattering length $a$.
This can also be expressed as $\lambda(T_0) \gg a$,
where $\lambda$ is the thermal wavelength
\begin {equation}
   \lambda(T) = \left(2\pi\hbar^2 \over m\kB T\right)^{1/2} .
\end {equation}
It is well known that, at distance scales large compared to the
scattering length $a$, an appropriate effective theory for a dilute Bose
gas is the second-quantized Schr\"odinger equation, together with
a chemical potential $\mu$ that couples to particle number density
$\psi^*\psi$, and
a $|\psi|^4$ contact interaction that reproduces low-energy scattering.
The corresponding Lagrangian is
\begin {equation}
   {\cal L} = \psi^* \left(
        - {i\hbar} \, \partial_t - {\hbar^2\over 2m} \, \nabla^2
        - \mu \right) \psi
    + {2\pi\hbar^2 a\over m} \, (\psi^* \psi)^2 .
\label {eq:L1}
\end {equation}
In this context the corresponding mean-field equation of motion is
called the Gross-Pitaevskii \  equation.%
\footnote{
   For a review, see \cite{Pitaevski}.
}
As with any effective theory, there are corrections represented by
higher-dimensional, irrelevant interactions (in the sense of the
renormalization group),%
\footnote{
   At short distances, the $\partial_t$ and $\nabla^2$ terms of the action
   $\int dt \> d^3x \> {\cal L}$ determine that
   times scales as (length)$^2$ and the scaling dimension of $\psi$ is
   (length)$^{-3/2}$.
}
such as $(\psi^*\psi)^3$ and
$\psi^* \nabla^4 \psi$.
However, higher and higher dimension operators are parametrically
less and less important if the distance scales
of interest are large compared to the characteristic scales ($a$)
of the atomic interactions.
The $(\psi^*\psi)^2$ term in the Lagrangian (\ref{eq:L1}) is in fact the
lowest-dimension irrelevant interaction, and it is adequate for
computing the leading-order effects of interactions in the
diluteness expansion.%
\footnote{
   For a discussion of analyzing corrections in this language, see
   ref.\ \cite{eff}, which extended earlier work on corrections by
   refs. \cite{corrections1}.
   A similar discussion for Fermi gases may be found in
   ref.\ \cite{fermi}.
}

Now treat the system at finite temperature using Euclidean time formalism.
The field can then be decomposed into frequency modes with Matsubara
frequencies $\omega_n = 2\pi n \kB T/\hbar$.
At sufficiently large distance scales ($\gg \lambda$),
and small chemical potential ($|\mu| \ll \kB T$),
the $-(\hbar^2/2m) \nabla^2 - \mu$ terms in (\ref{eq:L1}) become
small compared to the $O(\hbar\omega_n)$
time derivative term, provided $n\not=0$.
The non-zero Matsubara frequency modes then decouple from the
dynamics, leaving behind an effective theory of only the zero-frequency
modes $\psi_0$.  Roughly,
\begin {equation}
   {1\over\hbar} \int_0^{\hbar\beta} dt \int d^3x \> {\cal L} \to
   \beta \int d^3x \> \left[ \psi_0^* \left(
        - {\hbar^2\over 2m} \, \nabla^2
        - \mu \right) \psi_0
    + {2\pi\hbar^2 a\over m} \, (\psi_0^* \psi_0)^2 \right] 
\label {eq:L2}
\end {equation}
with $\beta = 1/k_B T$.
In detail, the parameters of (\ref{eq:L2}) are renormalized by coupling
to the non-zero modes, and there are again corrections in the form
of irrelevant (and even marginal)
interactions.  However, these effects are all suppressed in
the dilute limit%
\footnote{
   For the 3-dimensional effective theory (\ref{eq:L2}), the short-distance
   scaling dimension of $\psi_0$ is (length)$^{-1/2}$, the
   $(\psi_0^* \psi_0)^2$ interaction is relevant, and a
   $(\psi_0^* \psi_0)^3$ interaction would be marginal.
   Even though marginal, this last interaction can be ignored at the
   order of interest in the diluteness expansion because it has a
   small coefficient.
   For example, consider the term that would arise directly from the
   presence of a correction $g_3 (\psi^* \psi)^3$ to the original
   Lagrangian (\ref{eq:L1}).  That would lead to a $g_3 (\psi_0^* \psi_0)^3$
   term in (\ref{eq:L2}) which, after rescaling, would become a
   term proportional to $(mg_3/\hbar^2\lambda^4) (\phi^2)^3$
   in (\ref{eq:O2}).  Since
   $\lambda \simeq \lambda(T_0) \propto n^{-1/3}$ at the transition,
   the coefficient of
   this term is high order in the  diluteness expansion in $n^{1/3}$.
   Similarly, an effective $(\psi_0^*\psi_0)^3$ term arising from
   the 4-point interactions $(\psi^*\psi)^2$ and from
   integrating out physics at the scale $\lambda$ (due, for example, to
   non-zero Matsubara modes)
   would give rise
   to a $(\phi^2)^3$ term in (\ref{eq:O2}) with coefficient proportional
   to $u^3 \lambda^3  \propto \lambda^{-3}$.
}
and do not affect the computation
of $\Delta \Tc/T_0$ at leading order in $a n^{1/3}$.

It is then convenient to write
$\psi_0 = \hbar^{-1}(m \kB T)^{1/2}(\phi_1+i\phi_2)$ so that the
effective action $S = H/T$ becomes a conventionally normalized
O(2) field theory:
\begin {equation}
   S = \int d^3x \> \left[{1\over2} \, |\grad\phi|^2 + {1\over2} \, r \phi^2
             + {u\over 4!} \, (\phi^2)^2 \right] ,
\label{eq:O2}
\end {equation}
where $\phi$ is understood to be a 2-component real
vector $(\phi_1,\phi_2)$ and
\begin {equation}
   r = - {2 m \mu\over\hbar^2} \,,
   \qquad
   u = {96 \pi^2 a\over \lambda^2} \,.
\end {equation}


\subsection {Review of $\Delta\Tc/T_0$}

Our effective theory depends on two as yet undetermined parameters---$r$ and
$u$, or equivalently $\mu$ and $T$.  One constraint comes from fixing
particle number density $n$:
\begin {mathletters}%
\label{eq:constraints}%
\begin {equation}
   n = \langle \psi^* \psi \rangle
     = {m \kB T\over\hbar^2} \langle\phi^2\rangle .
\label {eq:n0}
\end {equation}%
At the critical temperature, a second requirement is that the system
have infinite correlation length, which requires
\begin {equation}
   \xi^{-1} = r + \Pi(0) = 0 ,
\label {eq:xi}
\end {equation}%
\end {mathletters}%
where $\Pi(p)$ is the proper self-energy of the $\phi$ field.
The two equations (\ref{eq:constraints}) determine the two unknowns
$r$ and $u$, and hence $\Tc$.
As noted by Baym {\it et al.},
the density equation (\ref{eq:n0}) can be rewritten as
\begin {equation}
   {\hbar^2 n\over m \kB T}
   = \int_\p {1\over p^2 + r + \Pi(p)}
   = \int_\p {1\over p^2 + [\Pi(p)-\Pi(0)]} ,
\label {eq:n}
\end {equation}
where the last equality uses (\ref{eq:xi}).
Throughout this paper, we
will use the notational shorthand
\begin {equation}
   \int_\p \equiv \int {d^3p\over (2\pi)^3}
\end {equation}
for momentum integrals.
(Technically, $p$ is a wave number rather than a momentum,
but we will use conventional $\hbar{=}1$
nomenclature, even though we have not set $\hbar$ to 1.)

The expression (\ref{eq:n}) for the density is ultraviolet (UV)
divergent and so receives contributions from short distance scales
where the effective theory breaks down.  This could be handled by
appropriately regulating the effective theory and then perturbatively
correcting the UV contribution.  As pointed out by Baym {\it et al.},
it is simpler to instead
consider the difference $n - n_0(T)$, where $n_0(T)$ is the same
expression in the absence of interactions (i.e., with $\Pi$ set to zero):
\begin {equation}
   {\hbar^2 [n-n_0(T)]\over m \kB T}
   = \int_\p \left[{1\over p^2 + r + \Pi(p)} - {1\over p^2}\right]
   = \int_\p \left[{1\over p^2 + [\Pi(p)-\Pi(0)]} - {1\over p^2}\right] .
\label {eq:dn}
\end {equation}
$n_0(T)$ represents the density a {\it non}-interacting Bose gas
has if its transition temperature is $T$.  It is given by
inverting (\ref{eq:T0}):
\begin {equation}
   n_0(T) = {\zeta(3/2)\over\lambda^3(T)} \,.
\end {equation}
This formula cannot be derived directly in the effective theory (\ref{eq:O2}),
but the difference $n-n_0$ in (\ref{eq:dn}) is insensitive to the UV
and so can be.

The above constraints are entirely adequate to systematically
determine $\Delta\Tc$ in the large $N$ expansion, but there is a
convenient way to simplify the bookkeeping a bit.
Baym {\it et al.} give a simple argument that, to leading order in
the density expansion,
\begin{equation}
   {\Delta\Tc\over T_0} \simeq - {2\over3} \, {[n - n_0(\Tc)] \over n} \,,
\label {eq:usedn}
\end {equation}
where the factor of $2/3$ in (\ref{eq:dn}) arises from the relation
$T \propto n_0^{2/3}$.
Combining (\ref{eq:usedn}) with (\ref{eq:dn}),
we can summarize as
\begin {equation}
   {\Delta\Tc\over T_0} \simeq
     - {2m \kB T_0\over 3\hbar^2 n} 
     \int_\p \left[{1\over p^2 + r + \Pi(p)} - {1\over p^2}\right]
   =
     - {2m \kB T_0\over 3\hbar^2 n} 
     \int_\p \left[{1\over p^2 + [\Pi(p)-\Pi(0)]} - {1\over p^2}\right] .
\end {equation}
to leading order in $a n^{1/3}$.
It's also useful to rephrase this, again in terms of the fields $\phi$ of
the effective theory, as
\begin {equation}
   {\Delta\Tc\over T_0} \simeq
     - {2m \kB T_0\over 3\hbar^2 n} \Delta\langle\phi^2\rangle ,
\label {eq:Tceq}
\end {equation}
where
\begin {equation}
   \Delta \langle\phi^2\rangle \equiv
        \langle \phi^2 \rangle - \langle \phi^2 \rangle_{\Pi\to0} .
\end {equation}

Note that the problem of calculating $\Delta\langle\phi^2\rangle$
from the action (\ref{eq:O2}), subject to the constraint (\ref{eq:xi}), has
only one dimensionful scale in it: $u$.  The length scale of this problem,
which will be the length scale of the physics that determines $\Delta\Tc/T_0$,
is therefore
\begin {equation}
   u^{-1} \sim {\lambda^2\over a}
\end {equation}
by dimensional analysis.
In the dilute limit $\lambda(\Tc) \gg a$, this length scale is large
compared
to $\lambda$, which justifies use of the O(2) effective theory
(\ref{eq:O2}).


\section {Review of leading order in $1/N$}
\label {sec:LO}

We now review the leading large-$N$ calculation of
$\langle\phi^2\rangle - \langle\phi^2\rangle_{\Pi\to0}$, and hence
of $\Delta\Tc/T_0$, by Baym {\it et al.}
The details of our calculation are slightly different than theirs,
and we will introduce techniques needed to proceed to
higher order.
We start with the standard large-$N$ generalization of the
O(2) scalar field theory (\ref{eq:O2}) to
an $O(N)$ scalar theory: replace $\phi$ by an $N$-component vector
and treat $N u$ as fixed in the $N \to \infty$ limit.
The reader should keep in mind that $u$ is therefore order $1/N$.
Standard $N$ power counting of Feynman diagrams consists of
a power of $u \sim 1/N$ for each 4-point vertex and a power of $N$ for each
flavor trace.

\begin {figure}
\vbox{
   \begin {center}
      \epsfig{file=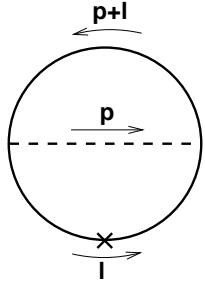,scale=.5}
   \end {center}
   \vspace*{-.1in}
   \caption{
       Diagrams contributing to $\Delta\langle\phi^2\rangle$ at leading
       order in $1/N$.
       \label{fig:LO}
   }
}
\end {figure}

\begin {figure}
\vbox{
   \begin {center}
      \epsfig{file=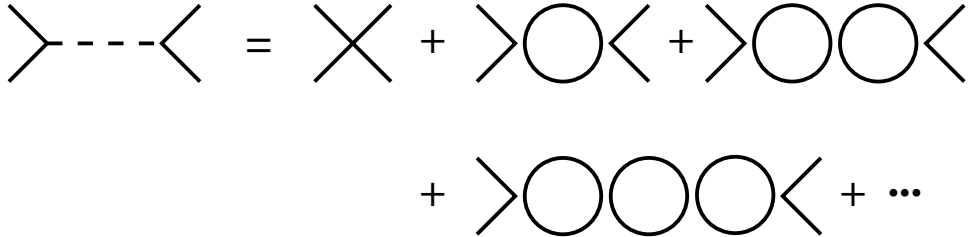,scale=.8}
   \end {center}
   \vspace*{-.1in}
   \caption{
       Bubble chains.  Unbroken lines denote flavor index contractions.
       \label{fig:chain}
   }
}
\end {figure}

\begin {figure}
\vbox{
   \begin {center}
      \epsfig{file=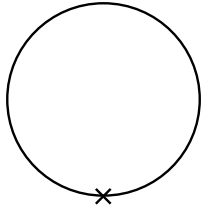,scale=.5}
   \end {center}
   \vspace*{-.1in}
   \caption{
       Diagram representing the non-interacting result
       $\langle\phi^2\rangle_{\Pi\to0}$.
       \label{fig:0}
   }
}
\end {figure}

The set of diagrams that determine
$\Delta\langle\phi^2\rangle$ at leading order in $1/N$
is depicted in Fig.\ \ref{fig:LO},
where the dashed line denotes bubble chains,
as shown in Fig.\ \ref{fig:chain}.
(For comparison, the diagram for $\langle\phi^2\rangle_{\Pi\to 0}$ is
shown in Fig.\ \ref{fig:0}.)
The cross denotes an insertion of the operator $\phi^2$, whose expectation
we are computing.
There is a simple way to summarize the effect on diagrammatic
perturbation theory of the $r \phi^2$ term
in the action (\ref{eq:O2}) and the constraint
(\ref{eq:xi}) that $r = -\Pi(0)$.

\begin {quote}
{\it Rule 1: Use massless (gapless)
scalar propagators $1/p^2$ when evaluating diagrams,
ignoring the $r\phi^2$ term in the action.
But whenever there is a one-particle irreducible sub-diagram $X$ that
represents a contribution to the $\phi$ proper self-energy $\Pi(p)$, then%
\footnote{
   This rule is unambiguous for calculating expectations such as
   $\langle\phi^2\rangle$.  It is potentially
   ambiguous for calculating the free
   energy---for example, a diagram like Fig.\ \ref{fig:LO} but without
   the cross on it.  In that case, it is ambiguous which sub-diagrams
   would be considered self-energy insertions.  A systematic way to
   treat the perturbation theory in all cases is to treat the $r \phi^2$
   term in the action as a perturbation, include it in Feynman diagrams
   as a 2-point vertex, and then set $r = -\Pi(0)$ order by order in
   perturbation theory.
}
replace $X(p)$ by $X(p)-X(0)$.}
\end {quote}

We note for later reference that, for the purpose of this rule,
a diagram that is cut in two pieces
only by cutting a single internal {\it dashed}\/ line
is still one-particle irreducible, because cutting the bubble chain
represented by a dashed line corresponds
to cutting two $\phi$ lines (Fig.\ \ref{fig:chain}).

The bubble chain sum shown in fig.\ \ref{fig:chain} is given by
\begin {equation}
   \raisebox{-17pt}{\mbox{\epsfig{file=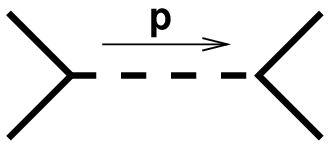}}}
         = - N^{-1} \f_p ,
\label {eq:chain}
\end {equation}
where
\begin {mathletters}
\label {eq:fp}
\begin {equation}
\label{eq:bubblesum}
   \f_p \equiv {1\over {3\over Nu} + \tilde \Sigma_0(p)}
\end {equation}
and $\tilde \Sigma_0(p)$ represents the basic massless bubble integral
(not summed over flavors)
\begin {equation}
   \tilde \Sigma_0(p) \equiv {1\over2} \int_\l {1\over l^2 |\l+\p|^2} \,.
\end {equation}%
\end {mathletters}%
In $d=3$ dimensions,
\begin {equation}
   \tilde \Sigma_0(p) = {1\over 16 p} \,.
\end {equation}
Putting everything together, the diagram of fig.\ \ref{fig:LO} gives
\begin {equation}
   \Delta\langle\phi^2\rangle =
   -
   \int_{\l\p} {\f_p\over l^4} \left[{1\over|\l+\p|^2} - {1\over p^2}\right]
   + O(N^{-1}) .
\label{eq:LOint}
\end {equation}

As pointed out by Baym {\it et al.} in \cite{baymN}, 
the above integral is not absolutely
convergent in three dimensions, and one must be careful to consistently
regulate the theory before proceeding.
Integrals that are not absolutely convergent are at best
ambiguous---they depend on the order one chooses to do the integrations.
For example, if one evaluates (\ref{eq:LOint}) directly in three dimensions,
doing the angular integrations first, then the $l$ integration, and
then the $p$ integration,
the result is zero.  This is not in fact the correct answer.
We will discuss this issue in some detail in order
to justify the correctness of our procedure for later evaluating
higher-order diagrams.

Baym {\it et al.}'s preferred method for the leading-order calculation
is to use dimensional regularization and evaluate everything in $d=3-\epsilon$
dimensions.
This is difficult at next order in $1/N$: the loop
integrals we shall encounter are sufficiently complicated
that evaluation in
$d=3-\epsilon$ dimensions seems hard.
Our strategy will be to instead always
reduce diagrams to well-defined three-dimensional integrals, which are
simpler to evaluate.  We imagine starting with some consistent
regularization scheme, like dimensional regularization, and will now
discuss how
to manipulate the integrals so that they will be absolutely convergent if
we set $d=3$.  We assume in what follows that the UV
regulator respects parity and is invariant under shifts $\p \to \p+\k$
of loop momenta $\p$.

Let's look at the divergences that cause absolute convergence of
the integral (\ref{eq:LOint}) to fail in three dimensions.   A simple one to
correct is the
behavior for $\l$ fixed and $p \to \infty$.  The large $p$
piece of the $\p$ integration then
behaves as $\int_\p \p\cdot\l/p^4$, which is logarithmically UV divergent
(from the point of view of absolute convergence).
This can be remedied by rewriting the {\it regulated}\/ version of
(\ref{eq:LOint}) by using the freedom to change the integration variable
$\p$ to $-\p$:
\begin {equation}
   \Delta\langle\phi^2\rangle_\LO =
   -
   \int_{\l\p} {\f_p\over l^4} \left[{1\over2|\l+\p|^2} + {1\over2|\l-\p|^2}
          - {1\over p^2}\right] .
\label{eq:LOint2}
\end {equation}
Now, if we throw away the UV regulator, the $\l$ fixed, large $p$
divergence is gone.  This sort of divergence is trivial, easy to remedy,
and won't have much practical impact on our calculations (given the
order in which we will eventually do integrations).
We'll simply acknowledge the issue in later calculations,
without emphasizing it, by writing (\ref{eq:LOint2}) as
\begin {equation}
   \Delta\langle\phi^2\rangle_\LO =
   -
   \int_{\l\p} {\f_p\over l^4} \left[{1\over|\l+\p|^2}
          - {1\over p^2}\right]_\pm ,
\label {eq:LOint2b}
\end {equation}
where the subscript $\pm$ means that one should average the expression with
$\p \to -\p$ (or equivalently with $\l \to -\l$).

Unfortunately, even (\ref{eq:LOint2}) is not
absolutely convergent.  There is still a logarithmic UV
divergence associated with
$\l$ and $\p$ {\it simultaneously}\/ becoming large ($l \sim p \to \infty$),
as can be seen by simple power counting and the fact that $\f_p$ approaches
a non-zero constant for large $p$.  Return to considering
(\ref{eq:LOint2}) with a UV regulator still in place.
We can eliminate the UV divergence by rewriting
\begin {equation}
   \Delta\langle\phi^2\rangle_\LO =
   -
   \int_{\l\p} {(\f_p-\f_\infty)\over l^4} \left[{1\over|\l+\p|^2}
          - {1\over p^2}\right]_\pm
   -
   \int_{\l\p} {\f_\infty\over l^4} \left[{1\over|\l+\p|^2}
          - {1\over p^2}\right]_\pm .
\end {equation}
The second integral vanishes, as can be seen by changing integration
variable $\p \to \p-\l$ in its first term.  So
\begin {equation}
   \Delta\langle\phi^2\rangle_\LO =
   -
   \int_{\l\p} {(\f_p-\f_\infty)\over l^4} \left[{1\over|\l+\p|^2}
          - {1\over p^2}\right]_\pm .
\label {eq:LOb}
\end {equation}
This is now UV convergent because $\f_p-\f_\infty \to 0$ as $p \to \infty$.
However, we have traded the logarithmic UV divergence for a
logarithmic infrared (IR) divergence,
associated with $p \sim l \to 0$.

We now need some sort of infrared regulator.
One physically motivated possibility for consistently regulating the infrared
would be to consider the system infinitesimally above the
critical temperature, so that all the massless scalar propagators
$1/p^2$ should be
replaced by massive ones
$1/(p^2 + M^2)$,
where the mass $M$ represents the inverse
correlation length $\xi^{-1}$.
This defines an absolutely convergent
integral in 3 dimensions, and
the limit $M \to 0$ would be taken only after
the integrations.

Massless propagators $1/p^2$ will be much easier to deal with, however,
in higher-order
calculations.  As a practical matter for computing diagrams, we
prefer to introduce as few massive propagators as possible.
It would be convenient, for example, to IR regulate (\ref{eq:LOb})
by introducing $M$ only in the $1/l^2$ propagators:
\begin {equation}
   \Delta\langle\phi^2\rangle_\LO =
   - \lim_{M\to 0}
   \int_{\l\p} {(\f_p-\f_\infty)\over (l^2+M^2)^2} \left[{1\over|\l+\p|^2}
          - {1\over p^2}\right]_\pm ,
\label {eq:LOreg}
\end {equation}
where $\f_p$ is still defined in terms of the massless bubble integral, as
in (\ref{eq:fp}).  One might worry that an {\it ad hoc}\/ procedure
of putting masses only on some propagators could be inconsistent, so let us
argue more carefully.
Return to the UV regulated version of (\ref{eq:LOint2})
and note that the integral is not sensitive to the the region of
integration where $l$ is infinitesimal, because this particular
integral is IR
convergent.  There's then no reason we can't modify the infrared behavior
of the integrand for infinitesimal $l$, without affecting the integral.
So, for instance,
\begin {equation}
   \Delta\langle\phi^2\rangle_\LO =
   - \lim_{M\to 0}
   \int_{\l\p} {\f_p\over (l^2+M^2)^2} \left[{1\over|\l+\p|^2}
          - {1\over p^2}\right]_\pm .
\end {equation}
But now, again rewriting $\f_p = (\f_p-\f_\infty)+\f_\infty$, the same steps
as before reproduce (\ref{eq:LOreg}).

Now that we have an absolutely convergent integral (\ref{eq:LOreg}), we can
do the integration in three dimensions and in any order we choose.  It's
convenient to do the $\l$ integral first:
\begin {equation}
   \int_\l
     {1\over (l^2 +M^2)^2} \left[{1\over|\l+\p|^2} - {1\over p^2}\right]_\pm
   = {1\over 8\pi M (p^2+M^2)} - {1\over 8\pi M p^2}
   = - {M \over 8\pi p^2 (p^2+M^2)} \,.
\end {equation}
The ``$\pm$'' prescription makes no difference to this particular
integral, because the $\l$ integration by itself is completely convergent
without it.  Note that naively setting $M$ to zero at this stage would
give the incorrect, zero result mentioned earlier.
Instead, we have
\begin {equation}
   \Delta\langle\phi^2\rangle_\LO =
   \lim_{M\to 0}
   \int_\p (\f_p-\f_\infty) {M \over 8\pi p^2 (p^2+M^2)} .
\label{eq:plup}
\end {equation}
The overall factor of $M$ in the numerator
is canceled by a linear IR divergence in the $\p$
integration, which is cut off by $M$.

For small $M$, the integral
(\ref{eq:plup}) is dominated%
\footnote{
  Some readers may worry that the integral (\ref{eq:plup}) is
  dominated by arbitrarily small $p \sim M \to 0$.
  They may worry  because at sufficiently
  small momentum our
  perturbative propagators are no longer
  good approximations to the
  full propagators.  The {\it full}\/ scalar propagators, for example, actually
  scale like $1/l^{2+\eta}$ rather than $1/l^2$ at small
  $l$ ($\ll Nu$), where the critical exponent $\eta$ is $O(N^{-1})$.
  The difference becomes significant when
  $l \lesssim N u \exp(-\eta) = N u \exp[-O(1/N)]$.
  One might worry that the sensitivity of
  (\ref{eq:plup})
  to $p\to 0$ is a sign that naive large $N$ perturbation theory
  must break down.
  It is important to realize, in the present case, that this
  infrared sensitivity
  is simply an artifact of our mathematical manipulations on the
  infrared-safe expression (\ref{eq:LOreg}).  Regardless of whether one
  used some sort of infrared-improved propagator in (\ref{eq:LOreg}), that
  expression is not sensitive to far-infrared momenta.  It is sensitive to
  momenta $\gtrsim Nu$, for which there is nothing wrong with a large-$N$
  expansion based on perturbative propagators.
}
by $p \sim M$.  So, in the limit
of $M \to 0$, we can simplify the calculation slightly by replacing
$\f_p-\f_\infty$ by $\f_0-\f_\infty$.  So
\begin {equation}
   \Delta\langle\phi^2\rangle_\LO =
   (\f_0 - \f_\infty)
   \int_\p {M \over 8\pi p^2 (p^2+M^2)}
   = {\f_0 - \f_\infty \over 32\pi^2}
   = - {Nu \over 96\pi^2} \,.
\label {eq:LO}
\end {equation}
When combined with the formula (\ref{eq:Tceq}) for $\Delta\Tc$, this reproduce
Baym {\it et al.}'s leading large $N$ result (\ref{eq:TcLO}), in
which $N$ has been set to 2.


\section{Next order in $1/N$}
\label {sec:NLO}

\begin {figure}
\vbox{
   \begin {center}
      \epsfig{file=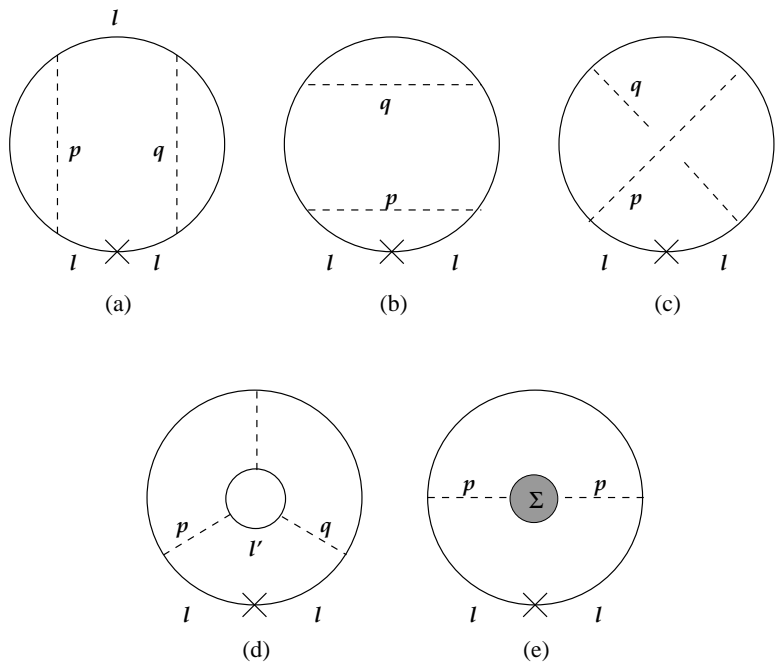,scale=.5}
   \end {center}
   \vspace*{-.1in}
   \caption{
       Next-to-leading order diagrams for $\Delta\langle\phi^2\rangle$.
       $\p,\q,\l,\l'$ label loop momenta, as used in the main text.
       \label{fig:NLO}
   }
}
\end {figure}

\begin {figure}
\vbox{
   \begin {center}
      \epsfig{file=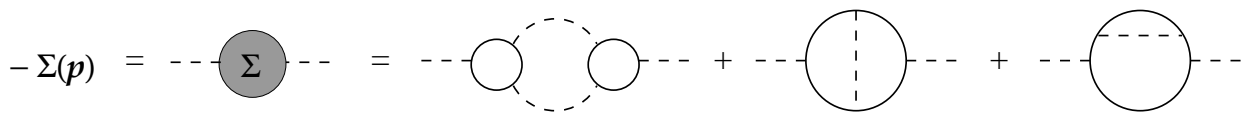,scale=.7}
   \end {center}
   \vspace*{-.1in}
   \caption{
       Diagrams for the $\sigma$ self-energy $\Sigma(p)$ at $O(N^0)$.
       \label{fig:Sigma}
   }
}
\end {figure}

\begin {figure}
\vbox{
   \begin {center}
      \epsfig{file=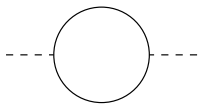,scale=.5}
   \end {center}
   \vspace*{-.1in}
   \caption{
       The one-loop bubble diagram.
       \label{fig:bubble}
   }
}
\end {figure}

\begin {figure}
\vbox{
   \begin {center}
      \epsfig{file=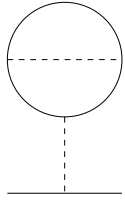,scale=.4}
   \end {center}
   \vspace*{-.1in}
   \caption{
       Example of a tadpole diagram.
       \label{fig:tadpole}
   }
}
\end {figure}

The diagrams which contribute to $\Delta\langle\phi^2\rangle$ at next order
in $1/N$ are shown in Figs.\ \ref{fig:NLO} and \ref{fig:Sigma}.
The diagrammatic expansion comes from the standard introduction of an
auxiliary field $\sigma$, represented by the dashed lines.%
\footnote{
   For a very quick review of standard large $N$, see, for example,
   section 2.1 of chapter 8 of \cite{coleman}.
   Some people might prefer to replace $\sigma$ by $i\sigma$ in the action
   (\ref{eq:Lsig1}),
   so that the Euclidean path integral for $\sigma$ is convergent,
   but it matters not at all for the purpose of large $N$ perturbation theory.
}
The $O(N)$ action of (\ref{eq:O2}) is rewritten as%
\begin {equation}
   S = \int d^3x \left[ {1\over2}\, | \Delta\phi|^2 + {1\over2}\, r \phi^2
       + {1\over 2}\, \phi^2 \sigma - {1\over 6u}\, \sigma^2 \right] .
\label {eq:Lsig1}
\end {equation}
The $\sigma$ propagator is then turned into the bubble chain of
fig.\ \ref{fig:chain} by resumming the basic massless bubble of
fig.\ \ref{fig:bubble} into the $\sigma$ propagator.  Technically, this is
accomplished by trivially rewriting (\ref{eq:Lsig1}) as
\begin {mathletters}
\label{eq:S}
\begin {eqnarray}
   S & = & S_0
       + S_{\rm subtractions}
       + \int d^3x \> \half \, \phi^2 \sigma ,\\
   S_0 & =& \int_\p \left[ \half \, \phi_{-\p} p^2 \phi_\p
             + \half \, \sigma_{-\p} (-N \f_{\p}^{-1}) \sigma_\p \right] ,\\
   S_{\rm subtractions} & = & \int_\p \left[ \half \, r \phi_{-\p} \phi_\p
     + \half \, \sigma_{-\p} \, N\tilde\Sigma_0(p) \, \sigma_\p \right] ,
\end {eqnarray}%
\end {mathletters}%
with $\f_p$ and $\tilde\Sigma_0$ given by (\ref{eq:fp}).
The terms designated $S_{\rm subtractions}$ may be ignored if one
follows the previous Rule 1 as well as
\begin {quote}
   {\it Rule 2: Do not include any diagrams that have the one-loop bubble,
   fig.\ \ref{fig:bubble}, as a sub-diagram.}
\end {quote}
Note that Rule 1 eliminates any tadpole sub-diagrams, such as
fig.\ \ref{fig:tadpole}.
Formal large $N$ counting of diagrams is simply to count a factor
of $N^{-1}$ for
each $\sigma$ propagator and a factor of $N$ for each $\phi$ loop.
The important momentum scale of the problem will be the scale
$p \sim Nu = O(N^0)$, where the $\sigma$ propagator (\ref{eq:chain})
makes the transition
from its small $p$ behavior ($\f_p \propto p$) to its large $p$ behavior
($\f_p \to \mbox{constant}$).
Some authors like to completely integrate $\phi$ out
of (\ref{eq:Lsig1}), but we prefer
to retain it, as there is then a more transparent relationship between
Feynman diagrams and the corresponding Feynman integrals.

In evaluating the diagrams of fig.\ \ref{fig:NLO}, we shall borrow techniques
from ref.\ \cite{wright}, where somewhat related diagrams were evaluated in
gauge theories with large numbers of scalars.  Our strategy will be to
do the $\phi$ loop integrals first, and then tackle the remaining integrals
associated with $\sigma$ propagators.


\subsection {Diagram a}

Let's start with Fig.\ \ref{fig:NLO}a.  The corresponding integral is
\begin {equation}
   \Delta\langle\phi^2\rangle_{\rm a}
   = N^{-1} \int_{\p\q} \f_p \f_q \int_\l {1\over l^6}
            \left[ {1\over |\l+\p|^2} - {1\over p^2} \right]_\pm
            \left[ {1\over |\l+\q|^2} - {1\over q^2} \right]_\pm .
\label {eq:a}
\end {equation}
As written, this integral is
absolutely convergent and can be evaluated, without
regularization, directly in three dimensions.
To do the $\l$ integration, however, we find it convenient to temporarily
introduce an IR regulator mass $M$.
We may then separately integrate
each of the terms of the integrand, which are not individually IR convergent.
We can also use $M$ as a trick for reducing powers of $l^{-2}$.
Specifically, we rewrite the $\l$ integral as the $M \to 0$ limit of
\begin {eqnarray}
&&
   \int_\l {1\over (l^2+M^2)^3}
            \left[ {1\over |\l+\p|^2} - {1\over p^2} \right]_\pm
            \left[ {1\over |\l+\q|^2} - {1\over q^2} \right]_\pm
\nonumber\\ && \qquad
   = {1\over2} \, {d^2\over d(M^2)^2} \,
   \int_\l {1\over (l^2+M^2)}
            \left[ {1\over |\l+\p|^2} - {1\over p^2} \right]_\pm
            \left[ {1\over |\l+\q|^2} - {1\over q^2} \right]_\pm
\nonumber\\ && \qquad
   = {1\over2} \, {d^2\over d(M^2)^2} \,
        \left[ I_1(\p,\q;M) - p^{-2} J_1(\q;M) - q^{-2} J_1(\p;M)
                   + p^{-2} q^{-2} \left(- {M \over 4\pi} \right) \right]_\pm ,
\nonumber\\ &&
\label {eq:linta}
\end {eqnarray}
where
\begin {equation}
   I_1(\p,\q;M) \equiv
   \int_\l {1\over (l^2+M^2) |\l+\p|^2 |\l+\q|^2} ,
\label {eq:I1def}
\end {equation}
\begin {equation}
   J_1(\p;M) \equiv \int_\l {1\over (l^2+M^2) |\l+\p|^2} ,
\label {eq:J1def}
\end {equation}
and%
\footnote{
   The integral $\int_\l (l^2+M^2)^{-1}$ is $-M/4\pi$ plus an $M$-independent
   UV divergence, and $M^2$ derivatives of the latter vanish.
   It's of course not necessary to introduce $\int_\l (l^2+M^2)^{-1}$ and this
   spurious UV divergence; one could simply evaluate
   $\int_\l (l^2+M^2)^{-3}$ directly.  But we find it
   convenient to consolidate the treatment of such integrals with
   that of the
   other terms in (\ref{eq:linta}).
}
\begin {equation}
   \int_\l {1\over (l^2+M^2)^3}
   = {1\over2} \, {d^2\over d(M^2)^2} \int_\l {1\over l^2+M^2}
   = {1\over2} \, {d^2\over d(M^2)^2} \left(- {M\over 4\pi}\right) .
\end {equation}

The integral $J_1$ is straightforward to evaluate.
A particularly simple way to evaluate $I_1$
is to make a conformal transformation which reduces it to the form of $J_1$.
The results of both integrals, and the conformal
transformation between them, are discussed in Appendix \ref{app:integrals}.
All we need here are the small $M$ expansions of those results, which
turn out to be
\begin {eqnarray}
   I_1(\p,\q;M) &=& {1\over 8 p q |\p-\q|} - {M \over 4\pi p^2 q^2}
                       - {M^2 \p\cdot\q\over 8 p^3 q^3 |\p-\q|}
\nonumber\\ && \hspace{5em}
                  + {M^3 (p^2+4\p\cdot\q+q^2) \over 12 \pi p^4 q^4}
                  + {M^4(3 (\p\cdot\q)^2 - p^2 q^2)\over 16 p^5 q^5 |\p-\q|}
                  + O(M^5) ,
\label {eq:I1expand}
\end {eqnarray}
\begin {equation}
   J_1(\p;M) = {1\over 8p} - {M\over 4 \pi p^2} + {M^3\over 12\pi p^4}
               + O(M^5) .
\end {equation}
Putting everything together,
\begin {eqnarray}
&&
   \int_\l {1\over (l^2+M^2)^3}
            \left[ {1\over |\l+\p|^2} - {1\over p^2} \right]_\pm
            \left[ {1\over |\l+\q|^2} - {1\over q^2} \right]_\pm
\nonumber\\ && \qquad
   = \left[{\hat\p\cdot\hat\q \over 8\pi M p^3 q^3}
            + {3(\hat\p\cdot\hat\q)^2 - 1 \over 16 p^3 q^3 |\p-\q|}\right]_\pm
            + O(M)
\nonumber\\ && \qquad
   = \left[{3(\hat\p\cdot\hat\q)^2 - 1 \over 16 p^3 q^3 |\p-\q|}\right]_\pm
            + O(M) .
\end {eqnarray}
We can now set $M = 0$.
All that will matter in the integral (\ref{eq:a}) is the average
$\langle \cdots \rangle_\theta$
over the angle between $\p$ and $\q$, which is
\begin {equation}
   \left\langle\int_\l {1\over l^6}
            \left[ {1\over |\l+\p|^2} - {1\over p^2} \right]_\pm
            \left[ {1\over |\l+\q|^2} - {1\over q^2} \right]_\pm
   \right\rangle_\theta
   = {1\over 40 p_>^6 p_<} ,
\end {equation}
where
\begin {equation}
  p_> \equiv {\rm max}(p,q),
  \qquad
  p_< \equiv {\rm min}(p,q).
\end {equation}
We are left with
\begin {equation}
   \Delta\langle\phi^2\rangle_{\rm a}
   = {1\over N} \int_{\p\q} {\f_p \f_q \over 40 p_>^6 p_<}
   = {1\over 2\pi^4 N} \int_0^\infty p^2 dp \int_0^p q^2 dq \>
         {\f_p \f_q \over 40 p^6 q} .
\end {equation}
The remaining integrals are easy to do, with the result
\begin {equation}
   \Delta\langle\phi^2\rangle_{\rm a}
   = {u \over 15 \pi^4} \left({\pi^2\over 6} - {5\over4}\right) .
\end {equation}


\subsection {Diagram b}

Fig.\ \ref{fig:NLO}b corresponds to
\begin {mathletters}
\begin {equation}
   \Delta\langle\phi^2\rangle_{\rm b}
   = N^{-1} \int_{\p\q} \f_p \f_q B_{\p\q} ,
\end {equation}
\begin {equation}
   B_{\p\q} \equiv
   \int_\l {1\over l^4} \left\{
         {1\over |\l+\p|^4} \left[{1\over|\l+\p+\q|^2} - {1\over q^2}\right]
         - {1\over p^4} \left[{1\over|\p+\q|^2} - {1\over q^2}\right]
     \right\}_\pm .
\end {equation}%
\end {mathletters}%
This contribution to $\Delta\langle\phi^2\rangle$ is again
absolutely convergent if the
subscript $\pm$ is taken to mean averaging over
$\p\to-\p$ and also over $\q\to-\q$.
It is convenient to now
rewrite the $\l$ integral as the $M_1, M_2 \to 0$ limit of
\begin {eqnarray}
&&
   \int_\l {1\over (l^2+M_1^2)^2}
     \left\{
         {1\over (|\l+\p|^2+M_2^2)^2}
              \left[{1\over|\l+\p+\q|^2} - {1\over q^2}\right]
         - {1\over (p^2+M_2^2)^2}
              \left[{1\over|\p+\q|^2} - {1\over q^2}\right]
     \right\}_\pm
\nonumber\\ && \qquad
   = {d\over d(M_1^2)} \, {d\over d(M_2^2)}
     \Biggl\{
        I_2(\p+\q,\q,M_1,M_2) - {1\over q^2} \, J_2(\p;M_1,M_2)
\nonumber\\ && \hspace{15em}
        - {1\over(p^2+M_2^2)} \left[{1\over|\p+\q|^2} - {1\over q^2}\right]
           \left(- {M_1\over 4\pi}\right)
     \Biggr\}_\pm ,
\nonumber\\ &&
\end {eqnarray}
where
\begin {equation}
   I_2(\p,\q;M_1,M_2) \equiv
   \int_\l {1\over l^2 (|\l+\p|^2+M_1^2) (|\l+\q|^2+M_2^2)} ,
\label {eq:I2def}
\end {equation}
\begin {equation}
   J_2(\p;M_1,M_2) \equiv \int_\l {1\over (l^2+M_1^2) (|\l+\p|^2+M_2^2)} .
\end {equation}
The results for $I_2$ and $J_2$, and their small $M_1,M_2$ expansions, are
given in Appendix \ref{app:integrals}.
The final result for the $\l$ integration, after taking the $M_1,\, M_2 \to 0$ 
limit, is
\begin {equation}
     B_{\p\q}
     = \left[{q - 2 p (\hat\p\cdot\hat\q) - 3 q (\hat\p\cdot\hat\q)^2
          \over 8 p^3 q^2 |\p+\q|^3} \right]_\pm ,
\end {equation}
with angular average
\begin {equation}
     \left\langle B_{\p\q} \right\rangle_\theta
     = {\theta(p-q)\over 4 p^6 q} \,,
\end {equation}
where $\theta(p-q)$ is the step function (1 for $p > q$; 0 for $p < q$).
The remaining integrals are easy to do, giving
\begin {equation}
   \Delta\langle\phi^2\rangle_{\rm b}
   = {u\over 3\pi^4} \left({\pi^2\over 6} - {5\over 4}\right) .
\end {equation}


\subsection {Diagram c}

Fig.\ \ref{fig:NLO}c can be evaluated as the others, but the final integrals
are a bit more complex.  The diagram gives
\begin {mathletters}
\begin {equation}
   \Delta\langle\phi^2\rangle_{\rm c}
   = N^{-1} \int_{\p\q} \f_p \f_q C_{\p\q} ,
\end {equation}
\begin {equation}
   C_{\p\q} \equiv
   \int_\l {1\over l^4} \left[
         {1\over |\l+\p|^2 |\l+\q|^2 |\l+\p+\q|^2}
         - {1\over p^2 q^2 |\p+\q|^2}
   \right]_\pm .
\end {equation}%
\end {mathletters}%
The $\l$ integration can be performed using methods similar to before:
\begin {equation}
   C_{\p\q} = - \lim_{M\to0} {d\over d(M^2)} \left[
        H(\p,\q;M) - {1\over p^2 q^2 |\p+\q|^2} \left(- {M \over 4\pi}\right)
   \right]_\pm ,
\end {equation}
\begin {equation}
   H(\p,\q;M) \equiv
      \int_\l {1\over (l^2+M^2) |\l+\p|^2 |\l+\q|^2 |\l+\p+\q|^2} \,.
\label{eq:H}
\end {equation}
$H$ can be reduced to the basic integrals $I_1$ and $J_1$ encountered
previously
by rewriting the numerator 1 in (\ref{eq:H}) as
\begin {equation}
   1 = {1\over 2\p\cdot\q + M^2} \left[
         (l^2+M^2) + |\l+\p+\q|^2 - |\l+\p|^2 - |\l+\q|^2
   \right]
\end {equation}
and then expanding the integrand into the corresponding four terms:
\begin {equation}
   H(\p,\q;M) = {1\over 2\p\cdot\q + M^2} \left[
      I_1(\p,\q;0) + I_1(\p,\q;M) - I_1(\p+\q,\q;M) - I_1(\p+\q,\p;M)
   \right] .
\end {equation}
Using the expansion (\ref{eq:I1expand}) of $I_1$, one obtains
\begin {equation}
   C_{\p\q} =
  \left [\frac{1}{16\, p^3\, q^3} 
  \left ( 1 + (\hat \p \cdot \hat \q)^{-2} \right ) 
  \left ( {1\over |\p - \q|} - {1\over |\p+\q|} \right ) +
  \frac{\hat \p \cdot \hat \q - (\hat \p \cdot \hat \q)^{-1}}
       {8\, p^2\, q^2\, |\p + \q|^3} \right ]_{\pm} .
\end{equation}
This simplifies, after applying the $[\dots]_\pm$ prescription, to
\begin{equation}
  C_{\p\q} = \frac{\hat \p \cdot \hat \q - (\hat \p \cdot \hat \q)^{-1}}
  {16\, p^2\, q^2}\, \left [\frac{1}{|\p + \q|^3} - \frac{1}{|\p - \q|^3} 
  \right ] .
\end{equation}
Angular averaging yields
\begin {equation}
   \langle C_{\p\q} \rangle_\theta
   = {1\over 8 p_>^7 x^2 (1+x^2)}
      \left[x + {\Sinh^{-1}x\over\sqrt{1+x^2}}\right]
   ,
\end {equation}
where
\begin {equation}
   x \equiv p_< / p_> .
\end {equation}
The remaining integrals over $\p$ and $\q$ are no longer so trivial.
Notice first that for an arbitrary function $f(x)$ one can rewrite
\begin {equation}
   {1\over N} \int_{\p\q}
     {\f_p \f_q \over p_>^7} f(x)
   = {1\over 2\pi^4 N} \int_0^\infty dp \int_0^1 dx 
     {\f_p \f_{xp} \over p^2} x^2 f(x)
   = - {8 u\over 3\pi^4} \int_0^1 dx {x^3 \ln x\over (1-x)} \, f(x) ,
\end {equation}
so that
\begin {equation}
   \Delta\langle\phi^2\rangle_{\rm c}
   = - {u\over 3\pi^4} \int_0^1 dx {x \ln x\over (1-x)}
      \left[{x\over 1+x^2} + {\Sinh^{-1}x\over(1+x^2)^{3/2}}\right] .
\end {equation}
This is easiest to evaluate numerically, giving
\begin {equation}
   \Delta\langle\phi^2\rangle_{\rm c}
   = {c u\over 3\pi^4},
\end {equation}
where $c \simeq 0.463715$.
We also have an analytic result:%
\footnote{
   Our inelegant, brute force method for obtaining this result is
   borrowed from a footnote of ref.\ \cite{wright}.  The hard part is the
   $\Sinh^{-1}$ term.  We change variables
   from $x$ to $y = x + \sqrt{1+x^2}$.  This turns $\Sinh^{-1}x$ into
   $\ln y$.  $\ln x$ can be written as a sum of terms of the form
   $\ln (y-a)$, and the change of integration variable makes the rest
   of the integrand a rational function of $y$.  We then split this rational
   function apart by partial fractions and do each integral, yielding
   di- and tri-logarithms of various arguments.  Finally, we
   use a zoo of polylogarithm identities \cite{polylogs} to simplify the
   answer.
}
\begin {equation}
   c =
       {\pi^2\over 48} \left[1 + {7\over\sqrt2}\,\ln(1+\sqrt2)\right]
       - {2\over3} \, L(3,\chi_8^{}) ,
\end {equation}
where
\begin{mathletters}
\label{eq:dirichletL}
\begin {equation}
   L(s,\chi_8^{}) =
   1 - {1\over \displaystyle{3^s}} - {1\over \displaystyle{5^s}}
     + {1\over \displaystyle{7^s}} + {1\over \displaystyle{9^s}}
     - {1\over \displaystyle{11^s}} - {1\over \displaystyle{13^s}}
     + {1\over \displaystyle{15^s}} + {1\over \displaystyle{17^s}} - \cdots
\end {equation}%
is a particular case of Dirichlet's $L$-function, and
\begin {equation}
   L(3,\chi_8^{}) = 0.958380454563\cdots .
\end {equation}%
\end{mathletters}


\subsection {Diagram d}

Fig.\ \ref{fig:NLO}d corresponds to
\begin {mathletters}
\begin {equation}
   \Delta\langle\phi^2\rangle_{\rm d}
   = - N^{-1} \int_{\p\q} \f_p \f_q \f_{|\q-\p|} D_{\p\q} ,
\end {equation}
\begin {eqnarray}
   D_{\p\q} &\equiv&
   \int_{\l'} {1\over {l'}^2 |\l'+\p|^2 |\l'+\q|^2}
   \int_\l {1\over l^4} \left[
         {1\over |\l+\p|^2 |\l+\q|^2}
         - \frac{1}{p^{2} q^{2}}
     \right]_{\pm\l}
\nonumber\\
   &=& - I_1(\p,\q;0) \lim_{M\to0} {d\over d(M^2)} \left[ I_1(\p,\q,M) -
             p^{-2} q^{-2} \left(- {M \over 4\pi}\right)
         \right]
\nonumber\\
   &=& {\hat\p\cdot\hat\q \over 64 p^3 q^3 |\p-\q|^2} \,.
\end {eqnarray}%
\end {mathletters}%
Here the subscript $\pm\l$ means we implicitly average over $\l \to -\l$
for absolute convergence.
Doing the remaining integrals by brute force, we find
\begin {equation}
   \Delta\langle\phi^2\rangle_{\rm d}
   = {u\over \pi^4} \left[{7\over12} \, \left(\zeta(3) - {\pi^2\over 6}\right)
               + {1\over6} \right] ,
\end {equation}
where $\zeta$ is the Riemann zeta function.
It's interesting to note that $\pi^2/6$ can also be written as $\zeta(2)$.


\subsection {Diagram e}

The final class of diagrams, fig.\ \ref{fig:NLO}e, correspond to the
leading order diagram, fig.\ \ref{fig:LO}, with the replacement
\begin {equation}
   -N^{-1}\, \f_\p \to (-N^{-1}\, \f_\p)[-\Sigma(p)](-N^{-1}\, \f_\p)
   \equiv -N^{-1}\, {\cal F}_\p ,
\label {eq:fpsub}
\end {equation}
where $\Sigma(p)$ represents the contribution to the $\sigma$ self-energy
at next-to-leading order, shown in the diagrams of fig.\ \ref{fig:Sigma}.
Making this substitution in the leading-order calculation (\ref{eq:LO})
gives
\begin {equation}
   \Delta\langle\phi^2\rangle_{\rm e} =
          { {\cal F}_0 - {\cal F}_\infty \over 32 \pi^2  } .
\label {eq:e}
\end {equation}
The nice feature of this relation is that we only need to calculate
$\Sigma(p)$ in the small and large $p$ limits.

The first self-energy diagram in fig.\ \ref{fig:Sigma} contributes
\begin {equation}
   \Sigma_1(\p) = - \half \int_{\q} \f_q \f_{|\p+\q|}
                  \int_{\l_1} {1\over l_1^2 |\l_1+\p|^2 |\l_1+\p+\q|^2}
                  \int_{\l_2} {1\over l_2^2 |\l_2+\p|^2 |\l_2+\p+\q|^2} .
\end {equation}
For small $\p$, the integration is dominated by $l_1, l_2 \sim p$ and
$q \sim Nu$.  So we can ignore $\l_1$, $\l_2$, and $\p$ compared to
$\q$ and write%
\footnote{
   For general $p$, the result is
   $N u \, \Sigma_1(p) = 24 \pi^{-2} z^{-3}
   {\rm Re}[2 \, \Li_2(-z) - 2 \,\Li_2(2+z) + \half\pi^2]$,
   where $z \equiv 48p/Nu$, and
   $\Li_2(z) \equiv -\int_0^z dx\>\ln(1-x)/x$
   is the dilogarithm function.
   One may double check that the $p\to0$ limit agrees with (\ref{eq:sig10}).
}
\begin {equation}
   \Sigma_1(\p) \to - \half \int_{\q} q^{-4} \f_q^2
                  \left[\int_{\l} {1\over l^2 |\l+\p|^2}\right]^2
                = - {Nu \over 48\pi^2 p^2} .
\label{eq:sig10}
\end {equation}
This diagram has a quadratic IR divergence for $\p = 0$.  In contrast,
the other two diagrams only diverge as {linear$\,\times\,$log} when $\p = 0$,
and so behave as $p^{-1} \ln p$ for small $p$.  In summary,
\begin {equation}
   \Sigma(\p) = - {Nu \over 48\pi^2 p^2} + O(p^{-1} \ln p)
\end {equation}
for small $\p$.  One may check by power counting diagrams and
sub-diagrams that $\Sigma(\infty) = 0$.
Then
\begin {equation}
   {\cal F}_0 = - {16\over 3\pi^2} \, u ,
   \qquad
   {\cal F}_\infty = 0 ,
\end {equation}
and%
\footnote{
   One may also check this answer by direct, brute-force calculation of all
   the diagrams associated with fig.\ \ref{fig:NLO}e.
}
\begin {equation}
   \Delta\langle\phi^2\rangle_{\rm e} =
          - {u \over 6 \pi^4 } .
\end {equation}

We conclude by mentioning one technical subtlety, glossed over above,
concerning absolute convergence.
The integration corresponding to the substitution
(\ref{eq:fpsub}) in the leading order analysis is
\begin {equation}
   \Delta\langle\phi^2\rangle_{\rm e} =
   -
   \int_{\l\p} {{\cal F}_p\over l^4} \left[{1\over|\l+\p|^2}
          - {1\over p^2}\right]_\pm .
\end {equation}
In contrast to the analogous leading-order expression (\ref{eq:LOint2b}),
this integral is not absolutely convergent in the infrared ($l \sim p \to 0$),
though it is convergent in the UV.  One might therefore worry about the
{\it ad hoc} introduction of an IR regulator $M$ in the calculation of this
graph.  However, this worry is easily bypassed by rewriting
\begin {equation}
   \Delta\langle\phi^2\rangle_{\rm e} =
   -
   \int_{\l\p} {({\cal F}_p-{\cal F}_0)\over l^4} \left[{1\over|\l+\p|^2}
          - {1\over p^2}\right]_\pm ,
\label{eq:eUV}
\end {equation}
which should be understood as regulated in the UV.  The UV-regulated
integral of the ${\cal F}_0$ factor vanishes.
Eq.\ (\ref{eq:eUV}) is now convergent in the IR, but logarithmically
divergent in the UV, just as the original leading-order integral
(\ref{eq:LOint2b}) was.
One can now follow through the same argument as in the leading-order case
to introduce an IR regulator and then remove the UV divergence, where
$\f_p$ in the leading-order analysis in now replaced by
${\cal F}_p - {\cal F}_0$.  The result is still (\ref{eq:e}).


\subsection* {Summary}

Summing all the diagrams then yields the total NLO contribution:
\begin {equation}
   \Delta\langle\phi^2\rangle_{\rm NLO}
   = {u\over 3\pi^4} \left[
           {7\over4} \, \zeta(3) - {3\over2} - {17\over240} \pi^2
           + {7\pi^2\over 48\sqrt2}\,\ln(1+\sqrt2) - {2\over3}\,L(3,\chi_8^{})
     \right] ,
\end {equation}
with $L(3,\chi_8^{})$ given by (\ref{eq:dirichletL}).
Combining with the leading-order result (\ref{eq:LO}),
\begin {equation}
   {\Delta\langle\phi^2\rangle_{\rm NLO} \over \Delta\langle\phi^2\rangle_\LO}
   = - {0.527198 \over N} \,,
\end {equation}
which is the relative NLO correction we presented for $\Tc$
in (\ref{eq:TcNLO}).


\section* {ACKNOWLEDGMENTS}

We are indebted to Eric Braaten for suggesting this project.
We are also grateful to Tim Newman and Genya Kolemeisky for useful discussions.
This work was supported, in part, by the U.S. Department
of Energy under Grant No.\ DE-FG02-97ER41027.

\appendix


\section {Basic Integrals}
\label {app:integrals}
Let's begin with the integral $J_2(\p;M_1,M_2)$.  This is quite easy to
do by standard methods (for example, by introducing a Feynman parameter),
and gives
\begin {eqnarray}
   J_2(\p;M_1,M_2) &\equiv& \int_\l {1\over (l^2+M_1^2) (|\l+\p|^2+M_2^2)}
\nonumber\\
   &=& {1\over 8\pi p} \,
       \Cos^{-1}\!\left( (M_1+M_2)^2 - p^2 \over (M_1+M_2)^2 + p^2\right)
\nonumber\\
   &=& {1\over 8p} - {(M_1+M_2) \over 4\pi p^2} + {(M_1+M_2)^3\over 12\pi p^4}
      + O(M^5) .
\end {eqnarray}
The integral $J_1(\p;m)$ of (\ref{eq:J1def}) is simply the special case
$J_1(\p;M) = J_2(\p;M,0)$.

The integral $I_2(\p,\q;M_1,M_2)$ of (\ref{eq:I2def})
can be related to $J_2(\p,\q;M_1,M_2)$
by generalizing a trick presented in ref.\ \cite{wright}.
The idea is to change integration variables from $\l$ to its conformal
inversion $\tilde \l \equiv \l/l^2$.  The
integration measure changes as
$(2\pi)^{-3} d^3l = (2\pi)^{-3} \tilde l^{-6} d^3\tilde l$.
Propagators can be written in terms of the new variable $\tilde\l$
as
\begin {eqnarray}
   {1\over l^2} &=& \tilde l^2 ,
\\
  {1\over|\l+\p|^2 + M^2} &=&
  {\tilde l^2 \over (p^2+m^2) [|\tilde\l+\tilde\P|^2 + \tilde M_p^2]} \,,
\end {eqnarray}
where
\begin {equation}
   \tilde\P \equiv {\p\over p^2 + M^2} \,,
   \qquad
   \tilde M_p \equiv {M \over p^2 + M^2} \,. 
\end {equation}
Making this change of variables,
\pagebreak  
\begin {eqnarray}
   I_2(\p,\q;M_1,M_2) &\equiv&
   {1\over (p^2+M_1^2) (q^2+M_2^2)}
   \int_{\tilde l}
        \left[ \left| \tilde\l + {\p\over p^2+M_1^2}\right|^2
                + \left(M_1\over p^2 + M_1^2\right)^2 \right]^{-1}
\nonumber\\ && \hspace{10em} \times
        \left[ \left| \tilde\l + {\q\over q^2+M_2^2}\right|^2
                + \left(M_2\over q^2 + M_2^2\right)^2 \right]^{-1}
\nonumber\\
   &=& {1\over (p^2+M_1^2) (q^2+M_2^2)} \,
       J_2\!\left[ {\p \over p^2+M_1^2} - {\q\over q^2+M_2^2} ;
             {M_1\over p^2+M_1^2} , {M_2\over q^2+M_2^2} \right] .
\end {eqnarray}
For the application of this paper, the relevant terms in the small $M_1,M_2$
expansion of $I_2$ are
\begin {eqnarray}
   I_2(\p,\q;M_1,M_2) &=&
   {1\over 8 p q |\p-\q|}
   - {1\over 4\pi |\p-\q|^2} \left({M_1\over p^2} + {M_2 \over q^2}\right)
   + O(M_1^2) + O(M_2^2)
\nonumber\\ && \hspace{3em}
   + {M_1 M_2\over 4\pi |\p-\q|^4}
           \left({M_1\over q^2} + {M_2 \over p^2}\right)
   + O(M_1^4) + O(M_2^4)
\nonumber\\ && \hspace{3em}
   + M_1^2 M_2^2 \,
         {[3 p q - 2 (p^2+q^2) (\hat\p\cdot\hat\q) 
             + p q (\hat\p\cdot\hat\q)^2] \over 8 p^2 q^2 |\p-\q|^5}
   + O(M^5) .
\end {eqnarray}

The integral $I_1$ of (\ref{eq:I1def}) is related by
$I_1(\p,\q;M) = I_2(\p,\p-\q;M,0)$ and gives
\begin {equation}
   I_1(\p,\q;M) =
   {\Cos^{-1}(2\omega_{\p\q}^2-1)\over
        8\pi M |\p-\q|^2\sqrt{\omega_{\p\q}^{-2}-1}} \,,
\end {equation}
where
\begin {equation}
   \omega_{\p\q} \equiv {M |\p-\q| \over \sqrt{(p^2+M^2)(q^2+M^2)}} \,.
\end {equation}
The small $M$ expansion is given in (\ref{eq:I1expand}).


\end {document}